\documentclass[10pt,aps,twocolumn,prd,superscriptaddress,showpacs,nofootinbib,noshowkeys,floatfix,preprintnumbers]{revtex4}
\usepackage{graphics,graphicx}
\usepackage{amsmath, amssymb}
\usepackage{multirow}
\usepackage{longtable}
\usepackage{color}

\usepackage[normalem]{ulem}  % \sout{old text} for strikeout

\begin{document}
\title{\bf Volume fluctuations and  higher order cumulants of the net baryon number}

\author{V. Skokov} \email[E-Mail:]{VSkokov@bnl.gov} \affiliation{%
  Physics Department, Brookhaven National Laboratory,
	Upton, NY 11973, USA
	}

\author{B.~Friman} \affiliation{%
  GSI Helmholtzzentrum f\"ur Schwerionenforschung, D-64291 Darmstadt,
  Germany}

\author{K.~Redlich} \affiliation{%
  Institute of Theoretical Physics, University of Wroclaw, PL--50204
  Wroc\l aw, Poland}


% \date{\today}

\begin{abstract}
We consider the effect of volume fluctuations on cumulants of the net baryon number.
Based on a general formalism, we derive universal expressions for the net baryon
number cumulants in the presence of volume fluctuations with an arbitrary probability distribution.
The relevance of these fluctuations for the baryon-number cumulants  and in particular for the ratios
of cumulants is assessed in the Polyakov loop extended quark-meson model  within the functional
renormalization group.  We show that the baryon number cumulants are generally enhanced by volume
fluctuations and that the critical behavior of higher order cumulants may be modified significantly.
\end{abstract}

\maketitle

\section{Introduction}
One of the goals of the experiments with ultrarelativistic heavy ion collisions
at SPS, RHIC and LHC energies is to probe the phase structure of strongly interacting matter and,
in particular, to identify the deconfinement and chiral symmetry restoration transitions. In this context,
the fluctuations of conserved charges may serve as a pertinent probe.

Fluctuations of the net baryon number and electric charge may provide an experimental signature for the hypothetical chiral critical endpoint~\cite{stp1,Stephanov:2011pb}. Moreover, as recently noted~\cite{karsch,karschr,pbm1,pbm2,sk1,Friman:2011pf,Skokov:2010uh,Skokov:2011rq,Mukherjee}, such fluctuations are also of interest at small baryon densities, since they reflect the critical dynamics of the underlying $O(4)$ transition, expected in QCD  in the limit of massless light quarks \cite{pisarski,karschl}.
Indeed, it was demonstrated that higher order cumulants change sign in the
crossover region of the QCD phase diagram~\cite{Skokov:2010uh,Friman:2011pf,Stephanov:2011pb}. Thus, the observation
of a strong suppression the higher order cumulants may  be used to identify  the chiral crossover transition in
experiment.

The  first measurements of fluctuations of the net  baryon number, more precisely of the net proton
number\footnote{As shown recently in Ref. \cite{Kitazawa:2012at}, the net nucleon number cumulants can, to a good approximation, be deduced from the measured net proton cumulants.},  in heavy ion collisions at RHIC were obtained  by the STAR
 Collaboration \cite{star}. The analysis of cumulants of the fluctuations and of the probability distributions confirmed, that the hadron resonance gas (HRG)  model, which yields a quantitative description of  particle yields in heavy ion collisions  \cite{hwa},   provides a useful reference for the non-critical background contribution to the  charge fluctuations \cite{karschr}. Thus, critical fluctuations related to the dynamics of the chiral transition should be reflected in deviations of the measured net charge fluctuations from the HRG baseline. In this context, higher order cumulants are of particular interest~\cite{Friman:2011pf,Skokov:2011rq}.

 A detailed analysis of experimental data on moments of net proton number fluctuations and their probability distributions  indeed exhibit deviations from the HRG.
To verify the origin of these deviations, one must identify and assess effects, unrelated to the critical dynamics, which can influence  the charge fluctuations.
 For instance, it was recently argued that constraints, owing
 to  the conservation of the total baryon number in nucleus-nucleus collisions \cite{volker} or
 experimental acceptances in terms of
kinematic variables \cite{starn},  might modify the non-critical background contributions
 to  higher order cumulants of the net proton number fluctuations.

In this paper we study volume fluctuations as a further possible source of non-critical fluctuations, not accounted for in the HRG model results. We first present a transparent derivation of the cumulants of net baryon number, including the effect of volume fluctuations. The resulting cumulants are expressed in terms of cumulants of the net baryon number distribution at fixed volume and cumulants of the probability distribution for volume fluctuations.

We also provide a more formal derivation, making use of the cumulant generating functions.
We stress that the final expressions are general, independent of the probability distributions for net baryon number and volume. The only assumption made is that the two sources of fluctuations are independent and
that  fluctuations of other
thermodynamic parameters are negligible. This assumption is most  likely justified for high energy heavy-ion collisions,
where the baryon chemical potential is close to zero. There, the thermalization\footnote{Thermalization is supported by the success of hydrodynamic and statistical models.}  results in
a freeze-out temperature independent of the initial conditions, while the volume
fluctuations are determined by the collision  geometry.
At lower  energies, fluctuations of the initial
temperature and chemical potential may take the system to different
freeze-out points,  along the freeze-out curve. Therefore, at lower energies, the fluctuations of temperature, chemical
potential and volume  are presumably correlated. Additional complications arise
if the system passes close to a possible critical end-point. In this case,   owing to the large correlation length, the volume fluctuations in the final state
may be  correlated with other thermodynamic variables.

With the  limitations discussed above, we focus on
the effect of volume fluctuations on the fluctuations of the net baryon number
in the vicinity of the chiral crossover transition at vanishing chemical potential.
We employ the functional renormalization group within the Polyakov loop
extended quark-meson model,  to properly account for the critical properties near the  chiral phase transition.

The paper is organized as follows: In the next section we obtain the corrections due to volume fluctuations  to the first four moments of the net baryon number fluctuations. In  Section III we derive a general expression for the corrected cumulants, valid to any order, obtained using the cumulant generating functions.
In Section IV we illustrate the role of volume fluctuations with a numerical study and finally in Section V we state our  conclusions.

\section{Heuristic approach}

Consider a fixed volume $V$, where
the net baryon number $B$ fluctuates with the probability distribution $P(B,V)$.
The n-th order  moments of the net baryon number
are then defined by
\begin{equation}
 \langle B^n \rangle_{V}  = \sum_{B=-\infty}^{\infty}B^n P(B,V).\label{eq2}
\end{equation}
It is convenient to introduce reduced cumulants, corresponding to the net baryon number fluctuations per unit volume. The first four reduced cumulants are
 \begin{align}\label{eq3}
 \kappa_1(T,\mu) &= {1\over V}\langle B \rangle_{V},\nonumber\\
 \kappa_2(T,\mu) &={1\over V} \langle (\delta B)^2 \rangle_{V},\nonumber\\
 \kappa_3(T,\mu) &= {1\over V}\langle (\delta B)^3 \rangle_{V},\\
 \kappa_4(T,\mu) &={1\over V} \left[ \langle (\delta B)^4 \rangle_{V} -3
 \langle (\delta B)^2 \rangle_{V}^2 \right], \nonumber
\end{align}
where
$\delta B = B-\bar B$ and $\bar{B}=\langle B\rangle_{V}$. The cumulants $\kappa_{i}$ are, to leading order, independent of the volume $V$. In the following we neglect subleading surface effects, which could lead to a residual volume dependence of the cumulants.

The volume dependence of the moments follows from (\ref{eq3}) and reads
\begin{align}\label{eq:inverse}
 \langle B \rangle_{V} &= \kappa_{1} V, \nonumber\\
 \langle B^{2} \rangle_{V}&= \kappa_{2} V  +\kappa_{1}^{2}\,V^{2},\nonumber\\
 \langle B^{3} \rangle_{V} &= \kappa_{3} V +  3 \kappa_{2}\kappa_{1}\,V^{2}+\kappa_{1}^{3}\,V^{3},\\
 \langle B^{4} \rangle_{V} &= \kappa_{4} V  +\left(4 \kappa_{3}\kappa_{1}+3\kappa_{2}^{2}\right) V^{2}+6\kappa_{2}\kappa_{1}^{2}\,V^{3}+\kappa_{1}^{4}\,V^{4}.\nonumber
 \end{align}
 The coefficients in Eq.~(\ref{eq:inverse}) are those of the Bell polynomials.

As an illustrative example, we consider the hadron resonance gas. In this model, the net baryon number fluctuations are given by the Skellam distribution~\cite{pbm1,pbm2} and the corresponding cumulants are particularly simple\footnote{The normalization of the generalized susceptibilities given in ~\cite{pbm2} differs from the cumulants used here by a factor $T^{3}$.}
\begin{equation}\label{HRG}
\kappa^{(HRG)}_{2n+1}=\frac{1}{V}(\bar B_{1}-\bar B_{-1}),\quad \kappa^{(HRG)}_{2n}=\frac{1}{V}(\bar B_{1}+\bar B_{-1}),
\end{equation}
where $\bar B_{1}=\langle B_{1}\rangle$ is the mean number of baryons  and
$\bar  B_{-1}= \langle B_{-1} \rangle$ that of anti-baryons in $V$. The corresponding moments are obtained by inserting the cumulants (\ref{HRG}) in (\ref{eq:inverse}).

We now allow for fluctuations of the volume. To this end, we introduce the
volume probability distribution  ${\cal P}(V)$, the corresponding moments
 \begin{align}\label{eq7}
 \langle {V^n}\rangle =\int V^n{\cal P}(V)dV,
\end{align}
and the reduced cumulants of the volume fluctuations, $v_n$. The latter  are  defined as in Eq. (\ref{eq3}) with the  replacements  $V\to\langle V\rangle$ and $B\to V$. Thus, e.g. $v_{1}=1$ and $v_{2}=(\langle V^{2}\rangle-\langle V \rangle^{2})/\langle V \rangle$.

In the presence of volume fluctuations the moments of the net baryon number  are given by
\begin{align}\label{eq8}
 \langle B^n \rangle  &= \int dV\,{\cal P}(V) \sum_{B=-\infty}^{\infty}B^n P(B,V)\nonumber\\
 &= \int dV\,{\cal P}(V) \langle B^{n} \rangle_{V}.
\end{align}
It is now straightforward to compute the reduced cumulants, including the effect of volume fluctuations. Using  Eqs. (\ref{eq3}), (\ref{eq:inverse}) and (\ref{eq8}), we find the general relations
\begin{eqnarray}
\label{eq9}
c_{1}&=&\kappa_{1}\nonumber\\
c_2 &=& \kappa _2 + \kappa _1^2 v_2, \nonumber\\
{c_3} &=& \kappa _3+3 \kappa _2 \kappa _1 v_2+ \kappa _1^3 v_3, \\
{c_4 }&=&\kappa _4 +(4 \kappa _3 \kappa _1+3 \kappa _2^2 )v_2+
6 \kappa _2 \kappa _1^2 v_3 + \kappa _1^4 v_4\nonumber ,
\end{eqnarray}
which are valid for arbitrary probability distributions, provided the fluctuations in baryon number and volume are independent. We note that the form of (\ref{eq9}) is determined by the volume dependence of the moments (\ref{eq:inverse}). Hence, the coefficients in (\ref{eq9}) are also given by the Bell polynomials.

\section{General derivation}
In the previous section, we explored the effect of volume fluctuations on the fluctuations of the net baryon number for the first few cumulants, where explicit calculations are tractable. In the following we derive a general expression for the cumulants, under the assumption that the fluctuations of baryon number and volume are independent.

\subsection{Formalism}
In general, the probability distributions introduced in section II are characterized by the corresponding cumulant generating
functions\footnote{We assume that the integrals in (\ref{chib}) and (\ref{chiv}) converge for $t$ and $s$ in an interval around the origin, so that the cumulant generating functions exist~\cite{lukacs}. 
%Moreover, we assume that the relevant cumulants exist.},
}
\begin{eqnarray}\label{chib}
\chi^B (t) &=& \ln \sum\limits_{B=-\infty}^{\infty}  P(B) \exp\left( B\,t\right), \\
\chi^V (s) &=& \ln \int\limits_{0}^{\infty} dV \, {\cal P}(V) \exp\left( Vs \right).
\label{chiv}
\end{eqnarray}
The cumulants are obtained by expanding $\chi^{B}$ and $\chi^{V}$ in a series about the origin.
The additivity of cumulants and thermodynamic principles imply,  that\footnote{  The  additivity of cumulants is valid only
 when  there are  no long range correlations in a system. This is the case near the  chiral crossover transtion, however  it is not applicable
 in the vicinity of a possible critical point.}
\begin{equation}
\chi^B(t) = V \cdot \zeta^B(t),
\label{zeta}
\end{equation}
where $\zeta^{B}$ is a volume-independent function. In fact, $\zeta^{B}$ is the generating function for the reduced cumulants, defined in Eqs. (\ref{eq3}):
\begin{equation}
\kappa_n=\left. \frac{d^n}{dt^n} \zeta^B(t) \right|_{t=0}.
\label{kappa}
\end{equation}
Similarly, we find for the reduced cumulants of volume fluctuations
\begin{equation}
v_n  = \frac {1}{\langle V \rangle}  \left.  \frac{d^n}{ds^n} \chi^V(s) \right|_{s=0}.
\label{v}
\end{equation}

Our aim is to compute cumulants of the net baryon number {\it including} the effects of volume fluctuations. These cumulants are obtained from the cumulant generating function
\begin{equation}
\phi^B (t) = \ln \int dV\, {\cal P}(V) \sum_B P(B,V) e^{Bt}.
\label{phi}
\end{equation}
Using Eq.~(\ref{zeta}) we find
\begin{equation}
\sum_B P(B,V) e^{Bt} = e^{V \zeta^B(t)},
\label{inv}
\end{equation}
and consequently
\begin{equation}
\phi^B(t) = \ln \int dV {\cal P} (V) e^{V \zeta^B(t)}.
\label{phi_inter}
\end{equation}

A comparison with the definition of  the cumulant generating function~(\ref{chiv}), yields
\begin{equation}
\phi^B(t) = \chi^V\left[ \zeta^B(t)  \right].
\label{final}
\end{equation}
This is the general form of the cumulant generating function for fluctuations of the net baryon number, including the effect of
volume fluctuations. The corresponding reduced cumulants are given by a Taylor expansion of $\phi^B(t)$  about  $t=0$,
\begin{equation}
c_n ={1\over {\langle V \rangle}} \left.  \frac{d^n}{dt^n} \phi^B(t) \right|_{t=0}.
\label{cn}
\end{equation}
We note   that since $\zeta^B(t=0)=0$,  no further normalization is needed in the calculation of the cumulants.

Using Fa\`{a} di Bruno's formula \cite{Faa},   we obtain a closed form expression for the cumulants,
\begin{equation}
c_n =     \sum\limits_{i=1}^{n} v_n\, B_{n,i}(\kappa_1, \kappa_2, \cdots, \kappa_{n-i+1}),
\label{cn_final}
\end{equation}
where $B_{n,i}$ are Bell polynomials. This equation confirms and extends our previous results for the first four cumulants, given in Eq.~(\ref{eq9}). Thus, for an arbitrary probability distribution for the fluctuations of net baryon number as well as for the fluctuations of the volume, Eq.~(\ref{cn_final}) yields cumulants that can be confronted with experiment. Conversely, given a model for the volume fluctuations, Eq.~(\ref{cn_final}) can be used to extract cumulants of the net baryon number in a fixed volume.

\subsection{Vanishing  chemical potential and  symmetric  volume fluctuations}

In the particular case  of vanishing  chemical potential,   all odd cumulants of net baryon number fluctuations
vanish, $\kappa_{2n+1}=0$. For the sake of simplicity, we also assume that the fluctuations of the volume are  symmetric, i.e.,
$v_{2n+1}=0$ for $n>1$.  In this case the first three non-vanishing cumulants are given by

\begin{eqnarray}
\label{Symm1}
c_2^{\rm s} &=& \kappa_2, \\
\label{Symm2}
c_4^{\rm s} &=& \kappa _4+3 \kappa _2^2 v_2  , \\
\label{Symm3}
c_6^{\rm s} &=& \kappa _6+15 \kappa _2 \kappa _4 v_2.
\end{eqnarray}
Thus, the cumulants $c_{n}^{s}$ for $n<8$ depend only on the second order cumulant of the volume fluctuations, $v_2$.
In other words, these cumulants are independent of the details of the probability distribution.

In the next section  we use the above  form to explore the effect of volume fluctuations on the cumulants of
net baryon number.

\section{Numerical results in the PQM model}
We  illustrate the influence  of
volume fluctuations on net baryon number fluctuations,  within a
model calculation. Of particular interest is the modification of higher order  moments
near the chiral crossover transition. We adopt the Polyakov loop-extended Quark
Meson Model (PQM) and compute the cumulants in a non-perturbative scheme, the functional renormalisation group. Details on the calculations and on the derivation of the net baryon number fluctuations
can be found in Ref.~\cite{Skokov:2010uh}.
In this exploratory calculation, we  consider only the case of
{\it symmetric} volume fluctuations and vanishing baryon chemical potential.
%For $\mu_B=0$,
%the cumulants $c_{n}$, with $n\geq 6$ are sensitive to the critical dynamics is the chiral transition.

%Gaussian fluctuations of the volume for a system at zero baryon chemical potential. In this case, all odd cumulants of the net baryon number vanish, i.e. $c_{2n+1}= \kappa_{2n+1}=0$.

In Ref.~\cite{Friman:2011pf} it was shown that near the chiral crossover transition, higher cumulants of the net baryon number ($n>4$) differ considerably from the predictions of the HRG model. In particular, it was suggested that negative values of $\kappa_6$ and $\kappa_8$ could be used to map out the chiral phase boundary.
This potential signal for the QCD phase transition may be affected by
volume fluctuations. Indeed, the second term
in Eq.~(\ref{Symm3}) yields a positive contribution to $c_6$. The strength of this contribution is directly proportional to the  second cumulant $v_2$ of volume fluctuations and may thus change the sign of $c_6$.
\begin{figure}
{\includegraphics[scale=0.4]{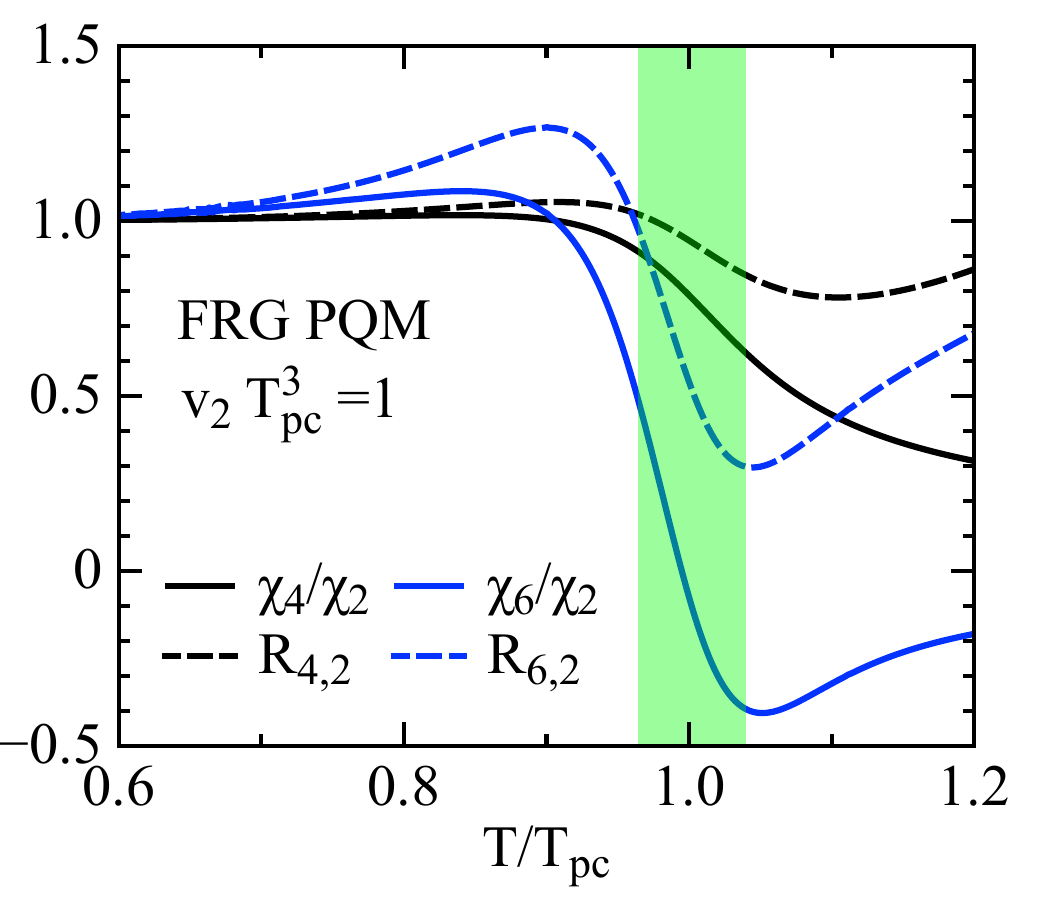}}
\caption{ The ratios $R_{4,2}$ and $R_{6,2}$ (defined in Eqs.~(\ref{Ratio42}, \ref{Ratios})) compared to $\chi_4/\chi_2$ and   $\chi_6/\chi_2$ (see Eq.~(\ref{R})) as functions of temperature, computed in the PQM model at vanishing chemical potential.    The probability distribution
for volume fluctuations is assumed to be symmetric, with the variance $v_2 T^3_{pc} = 1$, see the text for details.}
\label{r}
\end{figure}

%In what follows we show the illustrations for
%the value of  $v_2=0.6$. Because of the uncertainty we also demonstrate the dependence of the results as a function of $v_2$.
%employed in  Ref.~\cite{Kharzeev:2000ph}.

\begin{figure}
{\includegraphics[scale=0.4]{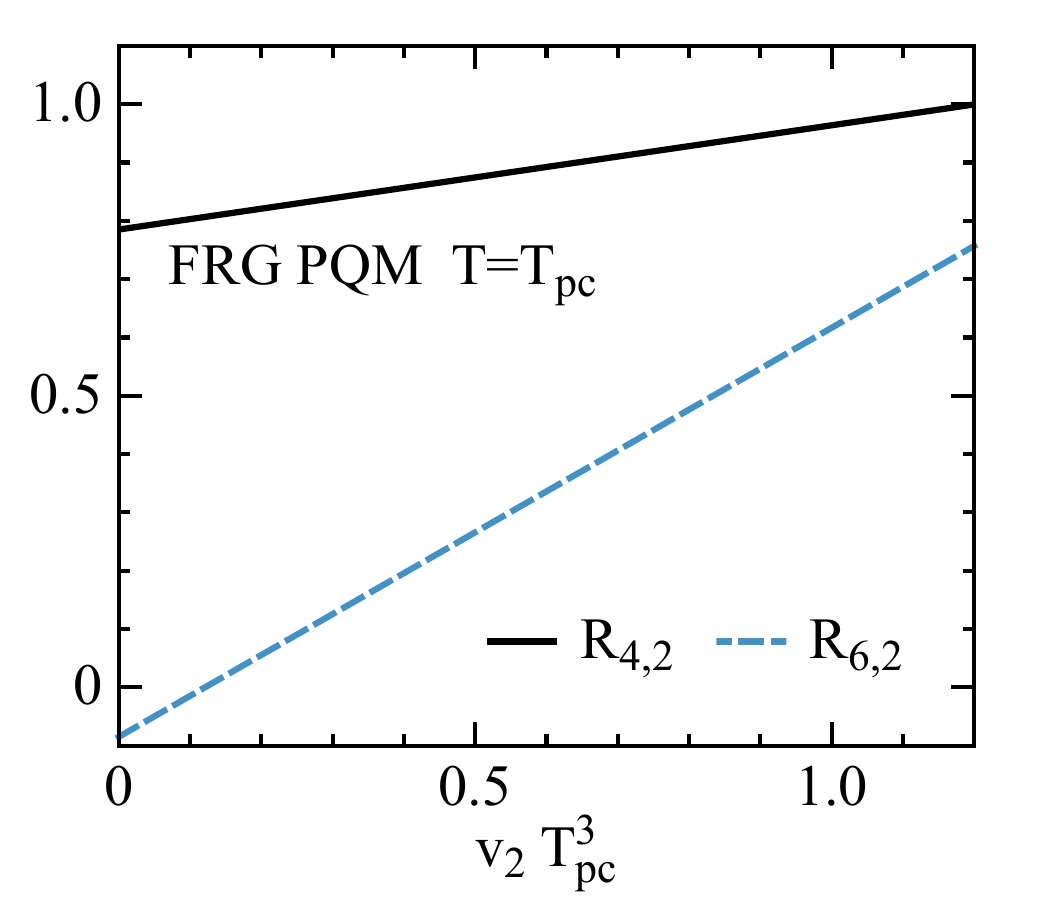}}
\caption{The ratios, $R_{4,2}$ and $R_{6,2}$ at the chiral crossover temperature $T_{\rm pc}$, obtained in the PQM model, as functions of the dimensionless variance of the volume fluctuations, $v_2 T_{\rm pc}^3$.}
\label{r_vs_v2}
\end{figure}

To proceed with the calculations in the PQM model, we relate the cumulants $\kappa_n$ to the
generalized susceptibilities $\chi_n$, defined by
\begin{equation}
\chi_n = \frac{\partial (p/T^4)}{\partial (\mu_B/T)^n} = \frac{\kappa_n}{T^3}.
\label{chin}
\end{equation}
It is useful to consider the ratios of cumulants,
\begin{equation}
R_{n,m} = \frac{c_n}{c_m},
\label{R}
\end{equation}
since many uncertainties cancel between the numerator and denominator.
Using (\ref{Symm2}) and (\ref{Symm3}) we thus find
\begin{eqnarray}
R_{4,2} &=& \frac{\chi_4}{\chi_2} + 3 \chi_2 \cdot  T^3 v_2 , \label{Ratio42}\\
R_{6,2} &=& \frac{\chi_6}{\chi_2} + 15 \chi_4 \cdot  T^3 v_2.
\label{Ratios}
\end{eqnarray}
%where $v_2 T^3 $ is  dimensionless.

Fig.~\ref{r}  shows the effect of volume fluctuations on the  $R_{4,2}$ and  $R_{6,2}$ ratios, obtained in the PQM model at fixed $T^3 v_2=1$. The contribution of the
volume fluctuations to both ratios are positive, and  grows with temperature.
This effect is also illustrated in Fig.~\ref{r_vs_v2}, where the ratios $R_{n,m}$ at the crossover transition temperature are shown as functions of $v_{2} T_{\rm pc}^3$.

The above  results indicate that volume fluctuations tend to suppress the signature of the chiral transition in the cumulants of net baryon number. Here, the  ratio  $R_{6,2}$ seems to be particularly sensitive.   Consequently,
the usefulness of fluctuations of conserved charges as a probe of criticality
in heavy ion collisions, depends crucially on the possibility to control volume fluctuations.

In general, volume fluctuations  are
difficult to assess. In heavy ion collisions  they  depend  on the centrality of the collision, on the  definition
used to fix the number of participants  and  on the  kinematic window, where the fluctuations are measured.
Thus,  $v_2$ is specific to a given experimental setup.

\begin{figure}
{\includegraphics[scale=0.4]{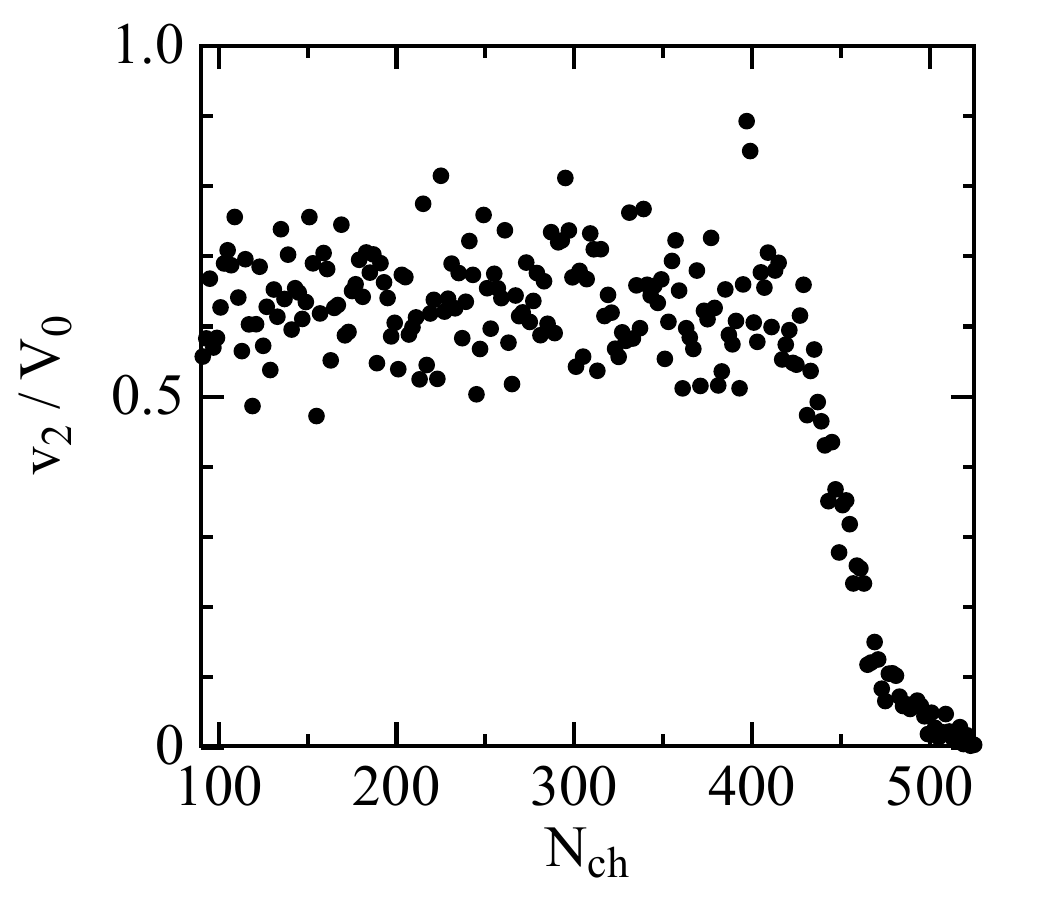}}
\caption{
The reduced variance of the volume fluctuations, $v_2$, as a function of the number of charged particles, $N_{\rm ch}$.
}
\label{glauber_v2}
\end{figure}

In order to explore the dependence of $v_2$ on the collision geometry, we performed a Glauber
Monte Carlo simulation, using the standard parameters for Au-Au collisions~\cite{Alver:2008aq}.
%In Fig.~\ref{glauber_prob} we
%show the probability distribution for number of participants in a given range of number of charged particles.
%To get information about volume fluctuations,
We assume that the volume is proportional to number of participants
$N_{\rm part}$ times a volume factor $V_0$, which we
fix to be equal to the volume of the proton, $V_{0}=2.83$ fm$^{3}$.
We compute the fluctuations in $N_{\rm part}$ for a fixed number of charged particles $N_{\rm ch}$.
A similar procedure is adopted by the
STAR collaboration
in their data analysis 
Ref.~\cite{star}. Since we adopted a small value for $V_{0}$, we expect that the resulting
estimate of $v_{2}$ is effectively a lower limit. In Fig.~\ref{glauber_v2}, we show the dependence of $v_2$ on the number of charged particles,
$N_{\rm ch}$. We find that the reduced variance, $v_2$, is approximately constant except for very central collisions, where the volume fluctuations
are strongly suppressed.

\begin{figure}
{\includegraphics[scale=0.4]{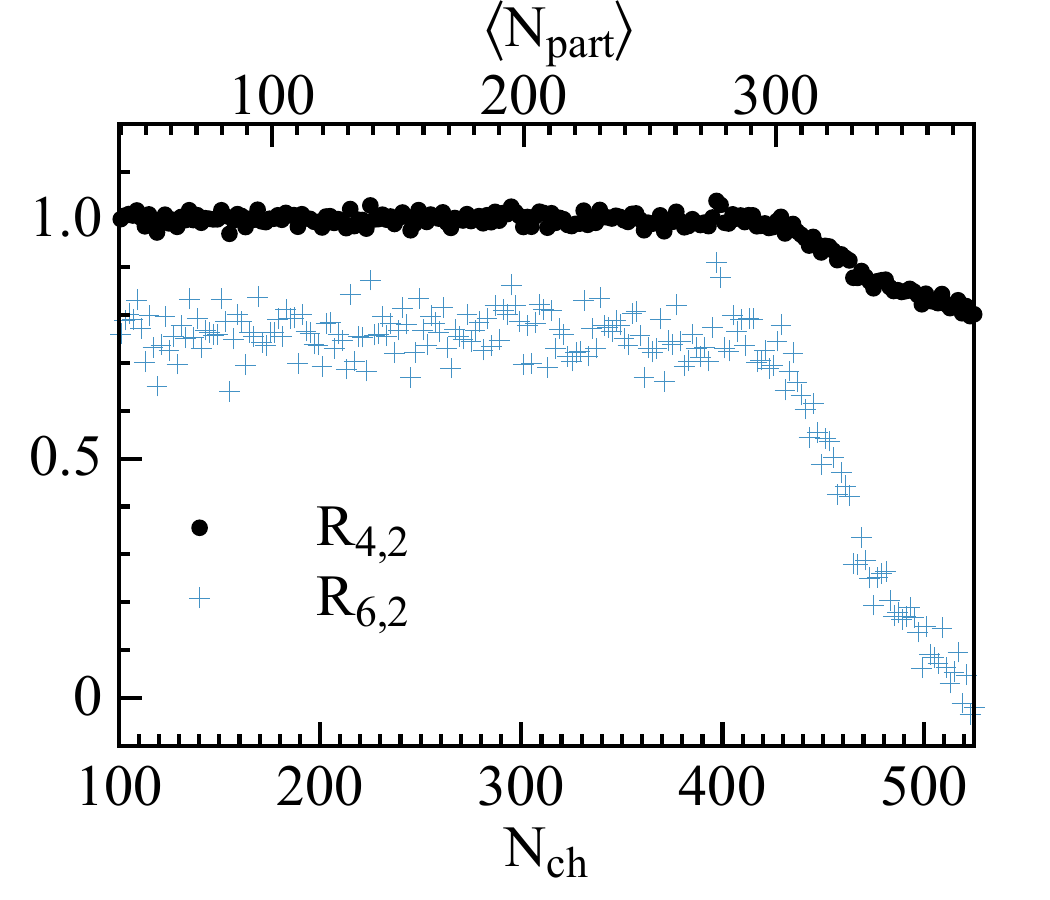}}
\caption{
Ratios of cumulants, $R_{4,2}$ and $R_{6,2}$, as functions of the number of charges particles in Au-Au collisions, based on the PQM model (see text for details). The freeze-out temperature is assumed to be equal to $T_{\rm pc}$.
}
\label{glauber_ratios}
\end{figure}

To assess the expected centrality dependence of the baryon number fluctuations, we use the Glauber result for $v_{2}$ and
assume that the freeze-out temperature $T_{\rm fr}=T_{\rm pc}$ and that it depends only weakly on centrality.
The resulting ratios $R_{4,2}$ and $R_{6,2}$ are shown in Fig.~\ref{glauber_ratios} as functions of $N_{\rm ch}$ and
$\langle N_{\rm part}\rangle$.

Recently, preliminary data on $R_{4,2}$ and $R_{6,2}$, obtained by the STAR collaboration, were reported in Refs.~\cite{Lizhu} and \cite{Luo:2011ts}.
It is found that both ratios are essentially independent
of the number of participants in collisions ranging from $\langle N_{\rm part} \rangle =2$ to  $350$.
As shown in Fig.~\ref{glauber_ratios},
our model also yields a weak dependence of $R_{4,2}$ and  $R_{6,2}$ on the number of participants, except
for very central collisions where volume fluctuations are suppressed.
We stress, however,  that this schematic model is not expected to yield a quantitative description of the experimental data.

%\begin{figure}
%{\includegraphics[scale=0.4]{glauber_prob}}
%\caption{
%The probability distribution for number of participants, $N_{\rm part}$, at given number of charged particles, $N_{\rm ch}$ for AuAu collisions at $\sqrt{s}=200$ GeV.
%}
%\label{glauber_prob}
%\end{figure}

\section{Conclusions}
We have studied  the influence  of  volume fluctuations on the  properties of  cumulants of net charge distributions  in heavy ion collisions.
 In particular, we have computed the contribution of volume fluctuations to ratios of net-baryon-number cumulants.
%\begin{figure}[t]
%{\includegraphics[scale=0.28]{r62_zeromu}}
%\caption{  Same as  Fig.~\ref{r42}, but for  $R_{6,2}$.}
%\label{r62}
%\end{figure}

In a heuristic approach we showed  explicitly how the corrections due to volume fluctuations arise. The resulting expressions, which hold for arbitrary probability distributions, were confirmed and extended in a general formalism, where we employed cumulant generating functions to obtain a closed form for the cumulants, including the effect of volume fluctuations.

We assessed the effect of volume fluctuations on the kurtosis $R_{4,2}$ as well as on ratios involving higher order cumulants, {\em viz.} $R_{6,2}$,  in the Polyakov loop extended quark-meson model.
A non-perturbative treatment of fluctuations was obtained by employing the functional renormalization
group.  We focused on the  structure of ratios of cumulants near  the chiral crossover transition, assuming that the probability distribution of volume fluctuations is approximately symmetric.

Finally, we showed  that phenomenologically relevant ratios of cumulants
of the net baryon number are enhanced by volume fluctuations. Consequently,  the
structure of these ratios, may be
significantly modified by volume fluctuations. Therefore, we conclude that  fluctuations of conserved charges in heavy ion collisions can provide
robust probes of  the chiral phase boundary if a good control of volume fluctuations can be achieved.

\pagebreak
\section*{Acknowledgments}
We thank P. Braun-Munzinger and A. Bzdak for stimulating discussions.
The research of V.S.\ was supported  under Contract No. DE-AC02-98CH10886 with the U. S. Department of Energy. The authors are supported in part by  the ExtreMe Matter Institute EMMI.
K.R.\  acknowledges partial support by the  National Science Centre (NCN).

%%%%%%%%%%%%%%%%%%%%%%%%%%%%%%%%%%%%%%%%%%%%%%%%

%\begin{thebibliography}{99}


\begin{thebibliography}{15}

\bibitem{stp1}
M.~A.~Stephanov, K.~Rajagopal and E.~V.~Shuryak,
  %``Signatures of the tricritical point in {QCD},''
	  Phys.\ Rev.\ Lett.\  {\bf 81}, 4816 (1998); Phys.\ Rev.\  D {\bf 60}, 114028 (1999).

\bibitem{Stephanov:2011pb}
  M.~A.~Stephanov,
	%``On the sign of kurtosis near the QCD critical point,''
	Phys.\ Rev.\ Lett.\  {\bf 107}, 052301 (2011).
\bibitem{karsch}
S.~Ejiri, F.~Karsch and K.~Redlich,
%``Hadronic fluctuations at the QCD phase transition,''
Phys.\ Lett.\  {\bf B633}, 275 (2006).
\bibitem{karschr}
 F.~Karsch and K.~Redlich,
  %``Probing freeze-out conditions in heavy ion collisions with moments of charge fluctuations,''
  Phys.\ Lett.\ B {\bf 695}, 136 (2011).
 \bibitem{pbm1}
 P.~Braun-Munzinger, B.~Friman, F.~Karsch, K.~Redlich and V.~Skokov,
  %``Net-proton probability distribution in heavy ion collisions,''
  Phys.\ Rev.\ C {\bf 84}, 064911 (2011).
 \bibitem{pbm2}
 P.~Braun-Munzinger, B.~Friman, F.~Karsch, K.~Redlich and V.~Skokov,
  %``Net-charge probability distributions in heavy ion collisions at chemical freeze-out,''
  Nucl.\ Phys.\ A {\bf 880}, 48 (2012).

\bibitem{sk1}
 V.~Skokov, B.~Friman, F.~Karsch and K.~Redlich,
  %``Charge fluctuations in chiral models and the QCD phase transition,''
  J.\ Phys.\ G  {\bf 38}, 124102 (2011).
\bibitem{Friman:2011pf}
 B.~Friman, F.~Karsch, K.~Redlich and V.~Skokov,
  %``Fluctuations as probe of the QCD phase transition and freeze-out in heavy ion collisions at LHC and RHIC,''
  Eur.\ Phys.\ J.\ C {\bf 71}, 1694 (2011).

%\cite{Skokov:2010uh}
\bibitem{Skokov:2010uh}
  V.~Skokov, B.~Friman and K.~Redlich,
	%``Quark number fluctuations in the Polyakov loop-extended quark-meson model at finite baryon density,''
	Phys.\ Rev.\ C {\bf 83}, 054904 (2011).
	%%CITATION = ARXIV:1008.4570;%%

 %\cite{Skokov:2011rq}
\bibitem{Skokov:2011rq}
  V.~Skokov, B.~Friman and K.~Redlich,
  %``Non-perturbative dynamics and charge fluctuations in effective chiral models,''
  Phys.\ Lett.\ B {\bf 708}, 179 (2012).


\bibitem{Mukherjee}
F. Karsch, E. Laermann, C. Miao, S. Mukherjee, P. Petreczky, C. Schmidt,
W. Soeldner and W. Unger,
%``The phase boundary for the chiral transition in (2+1)-flavor QCD at small values of the chemical potential,''
Phys. Rev. D {\bf 83}, 014504 (2011).

\bibitem{pisarski}
R.~D.~Pisarski and F.~Wilczek,
  %``Remarks on the Chiral Phase Transition in Chromodynamics,''
  Phys.\ Rev.\ D {\bf 29}, 338 (1984).
\bibitem{karschl}	
S. Ejiri, et al., Phys. Rev. D {\bf 80},  094505 (2009).
O. Kaczmarek, et al.,   Phys. Rev. D {\bf 83},   014504 (2011).

%\cite{Kitazawa:2012at}
\bibitem{Kitazawa:2012at}
  M.~Kitazawa and M.~Asakawa,
  %``Relation between baryon number fluctuations and experimentally observed proton number fluctuations in relativistic heavy ion collisions,''
  arXiv:1205.3292 [nucl-th].


\bibitem{star}
  M.~M.~Aggarwal {\it et al.} [STAR Collaboration],
  Phys.\ Rev.\ Lett.\  {\bf 105}, 022302 (2010).
  X.~Luo, {\it et al.}, [for the STAR Collaboration],
arXiv:1106.2926v1.

\bibitem{hwa}
P. Braun-Munzinger, K. Redlich, J. Stachel, in Quark-Gluon Plasma
3, Eds. R.C. Hwa and X.N. Wang, (World Scientific Publishing, 2004).
A. Andronic, P. Braun-Munzinger, and J. Stachel, Acta Phys. Polon. {\bf B40},
1005 (2009).

\bibitem{volker}
 A.~Bzdak, V.~Koch and V.~Skokov,
  %``Baryon number conservation and the cumulants of the net proton distribution,''
  arXiv:1203.4529 [hep-ph].
\bibitem{starn}
 P.~Garg, D.~K.~Mishra, P.~K.~Netrakanti, B.~Mohanty, A.~K.~Mohanty, B.~K.~Singh and N.~Xu,
  %``Conserved number fluctuations in a hadron resonance gas model,''
  arXiv:1304.7133 [nucl-ex].

\bibitem{lukacs} E. Lukacs, Characteristic functions, (Griffin, London, 1970).

\bibitem {Faa}
W.~P.~Johnson, Amer.\ Math.\ Monthly {\bf  109}, 217  (2002).

\bibitem {Lizhu}
Lizhu Chen, talk on Quark Matter 2012 International Conference.

%\cite{Luo:2011ts}
\bibitem{Luo:2011ts}
  X.~-F.~Luo [STAR Collaboration],
	  %``Probing the QCD Critical Point with Higher Moments of Net-proton Multiplicity Distributions,''
		  J.\ Phys.\ Conf.\ Ser.\  {\bf 316}, 012003 (2011)
			  [arXiv:1106.2926 [nucl-ex]].
				  %%CITATION = ARXIV:1106.2926;%%
					  %14 citations counted in INSPIRE as of 25 Mar 2013

%\cite{Alver:2008aq}
\bibitem{Alver:2008aq}
  B.~Alver, M.~Baker, C.~Loizides and P.~Steinberg,
	          %``The PHOBOS Glauber Monte Carlo,''
						                  arXiv:0805.4411 [nucl-ex].
															                          %%CITATION = ARXIV:0805.4411;%%
													                                  %91 citations counted in INSPIRE as of 22 Mar 2013

%%\cite{Kharzeev:2000ph}
%\bibitem{Kharzeev:2000ph}
%  D.~Kharzeev and M.~Nardi,
	%``Hadron production in nuclear collisions at RHIC and high density QCD,''
%	Phys.\ Lett.\ B {\bf 507}, 121 (2001).
	%%CITATION = NUCL-TH/0012025;%%



%\cite{Stephanov:2011pb}

	%%CITATION = ARXIV:1104.1627;%%

\end{thebibliography}
\end{document}